\documentclass[runningheads]{llncs}
\usepackage[T1]{fontenc}
\usepackage{graphicx}
\usepackage{multirow}
\usepackage{cite}
\usepackage{hyperref}
\usepackage{lscape}
\usepackage{dingbat}
\usepackage{color, colortbl}
\usepackage{algorithm}
\usepackage{algpseudocode}
\usepackage{amssymb}

\definecolor{TableGray}{gray}{0.9}
\definecolor{TableLightCyan}{rgb}{0.88,1,1}

\usepackage{color}

\begin{document}
%



%
\title{Advanced Tumor Segmentation in Medical Imaging: An Ensemble Approach for BraTS 2023 Adult Glioma and Pediatric Tumor Tasks}
\titlerunning{Advanced Tumor Segmentation in Medical Imaging}
%
\author{Fadillah Maani$^*$\inst{1} \and
Anees Ur Rehman Hashmi$^*$\inst{1} \and
Mariam Aljuboory\inst{1,2} \and
Numan Saeed\inst{1} \and
Ikboljon Sobirov\inst{1} \and
Mohammad Yaqub\inst{1}
}
\authorrunning{F. Maani et al.}
%
\institute{Mohamed bin Zayed University of Artificial Intelligence, Abu Dhabi, UAE
\email{\{fadillah.maani,anees.hashmi\}@mbzuai.ac.ae} \and
Northwestern University, Illinois, USA}
\maketitle              
\def\thefootnote{*}\footnotetext{Equal contribution}

\begin{abstract}
Automated segmentation proves to be a valuable tool in precisely detecting tumors within medical images. The accurate identification and segmentation of tumor types hold paramount importance in diagnosing, monitoring, and treating highly fatal brain tumors. The BraTS challenge serves as a platform for researchers to tackle this issue by participating in open challenges focused on tumor segmentation. This study outlines our methodology for segmenting tumors in the context of two distinct tasks from the BraTS 2023 challenge: Adult Glioma and Pediatric Tumors. Our approach leverages two encoder-decoder-based CNN models, namely SegResNet and MedNeXt, for segmenting three distinct subregions of tumors. We further introduce a set of robust postprocessing to improve the segmentation, especially for the newly introduced BraTS 2023 metrics. The specifics of our approach and comprehensive performance analyses are expounded upon in this work. \textit{Our proposed approach achieves third place in the} \texttt{BraTS 2023 Adult Glioma Segmentation Challenge} with an average of 0.8313 and 36.38 Dice and HD95 scores on the test set, respectively.

\end{abstract}

\section{Introduction}
 Cancerous brain tumors are one of the deadliest types of central nervous system tumors~\cite{alexander2017adult}, and they account for the highest number of cancer-related deaths in pediatrics. Glioma is a brain tumor that originates from glial cells, which provide structure and support to the nerve cells. Astrocytoma is the type of glioma that occurs in the astrocytes ( a type of glial cells), which are responsible for a healthy brain environment. The treatment options for brain tumors include surgery, chemotherapy, and radiation therapy. Neurologists, oncologists, and radiologists work together to develop the treatment plan for patients. Magnetic Resonance Imaging (MRI) scans provide information on the patient's internal structure, tissue, and organs. The scans are used for treatment plans and to assess the performance of the treatment~\cite{owrangi2018mri}.

Radiologists predict tumor classification and whereabouts from MRI scans. As a result of the shortage of healthcare workers in some countries, radiologists are often overworked and under constant pressure, which at times naturally leads to human error. Manual segmentation of the tumor can be very time-consuming for radiologists. Automatic segmentation can increase the accuracy of tumor classifications and improve the workload for radiologists by providing them with additional resources that can help them feel supported.

The annual Medical Image Computing and Computer Assisted Interventions (MICCAI) conference hosts medical imaging challenges for research teams to participate internationally. One of MICCAI’s challenges is BraTS \cite{menze2014multimodal}, the brain tumor segmentation that consists of nine tasks this year. Initially, BraTS started as a challenge that only aimed at adult glioma \cite{bakas2017advancing, bakas2017segmentationlgg, bakas10segmentationgbm}; however, recently, it has expanded its dataset to increase the diversity of the segmentation tasks. The additional population in the expanded dataset includes sub-Saharan African patients, pediatrics, and meningioma tumors. The aim of expanding the dataset is to account for different brain tumors, a diverse range of image quality, and tumor sizes. This paper focuses on the adult and pediatric datasets.

Artificial intelligence uses neural networks to train on pre-existing data to learn the boundaries of brain tumors. Automatic segmentation is made possible by deep learning models that analyze the MRI dataset. As a result, physicians can use the algorithms to effectively and accurately identify tumors. This paper highlights the use of MedNeXt and SegResNet models for fully automated brain tumor segmentation. The training and validation occurred simultaneously when testing the different models on the dataset. BraTS provided researchers with the ground truth to accurately assess the performance of the models based on the predictions. The tumor segmentations were outlined by radiologists and reviewed by neurologists to ensure the accuracy of the data.

Manual segmentations are often time-consuming; semi-automated segmentation is a computer-based model with human contributions. Generally, the expert radiologist needs to manually guide the algorithm by providing the outlines for the segmentations. Then, the model uses the information imputed and automatically segments the region of interest in scans. This technique was one of the initial transitions to artificial intelligence segmentations ~\cite{xie2005semi}. Two-stage segmentation frameworks are fully automatic but consist of two steps: balancing the classes and refining to the proper proportions. For brain tumors, there are several different imbalance classes; balancing the classes allows for equal representation and prevents biased training. Refining the predictions can improve the segmentation results by matching the accurate class distribution ~\cite{pereira2016brain}. While both the semi-automated and two-stage segmentations improve the workload in radiology, fully automated segmentations are the most efficient and consistent. 


Our main contributions are the following:
\begin{itemize}
    \item A proposed ensemble of deep learning models for adult and pediatric brain tumor segmentation from the BraTS 2023 challenge.
    \item An integration of the deep supervision component with the models and investigation of its effects on the performance. 
    \item A thorough investigation and analysis of post-processing techniques for the brain tumor segmentation tasks. 
\end{itemize}

\section{Methods}
\subsection{Dataset}

\noindent \textbf{BraTS-Adult Glioma.}
The dataset contains multi-institutional structural MRI scans of four different contrasts: pre and post-gadolinium T1-weighted (T1 and T1CE), T2-weighted (T2), and T2-weighted fluid-attenuated inversion recovery (T2-FLAIR). The dataset was acquired from multiple institutions with high variance in many aspects, including brain shape, appearance, and tumor morphology. %
The dataset \cite{baid2021rsna,karargyris2023federated} comprises 1251 training and 219 validation brain MRI scans. Furthermore, the final model performance will be evaluated on the testing set that will not be released.

\noindent \textbf{BraTS-PEDs.}
A total of 228 high-quality are acquired from 3 different institutions, including  Children’s Brain Tumor Network (CBTN), Boston’s Children Hospital, and Yale University. The acquired MRI modalities are T1, T1Gd, T2, and T2-FLAIR. All the included pediatric subjects contain histologically-approved high-grade glioma, i.e., high-grade astrocytoma and diffuse midline glioma (DMG), including radiologically or histologically-proven diffuse intrinsic pontine glioma (DIPG) \cite{kazerooni2023brain,karargyris2023federated}. The released training and validation datasets comprise 99 and 45 subjects, respectively.

\noindent \textbf{Segmentation labels.}
The provided annotations consist of the GD-enhancing tumor (ET), the peritumoral edematous or invaded tissue (ED), and the necrotic tumor core (NCR). Instead of being evaluated on these labels, segmentation performance is assessed based on the different glioma sub-regions: enhancing tumor, tumor core (NCR + ET), and whole tumor (NCR + ET + ED).



\noindent \textbf{Preprocessing.}
The provided MRI scans were preprocessed by co-registration of the four modalities to a standard SR124 template \cite{rohlfing2010sri24}, isotropic interpolating to meet $1mm^3$ resolution and skull-stripping. All of the MRI image sizes are uniform $240\times240\times155$. We further preprocess each scan by cropping the foreground, normalizing voxels with non-zero intensities, and finally stacking the four modalities into a single image. Yet, we experienced a bottleneck when we applied the preprocessing steps on the fly. Thus, we preprocess all MRI scans and store them in the \texttt{.npy} format, and then load the \textit{Numpy} arrays during training.

\begin{figure}[t!]
\centering
{\includegraphics[width=0.9\columnwidth]{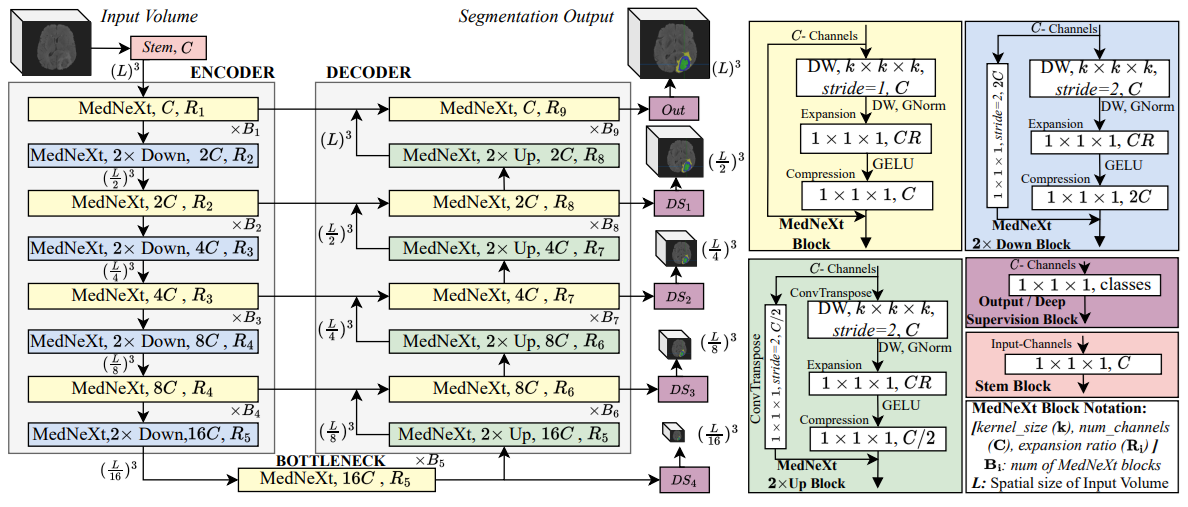}}

\caption{The MedNeXt network~\cite{roy2023mednext}. }
\label{fig:model}
\end{figure}

\subsection{Models}
\begin{figure}[t!]
\centering
{\includegraphics[width=0.9\columnwidth]{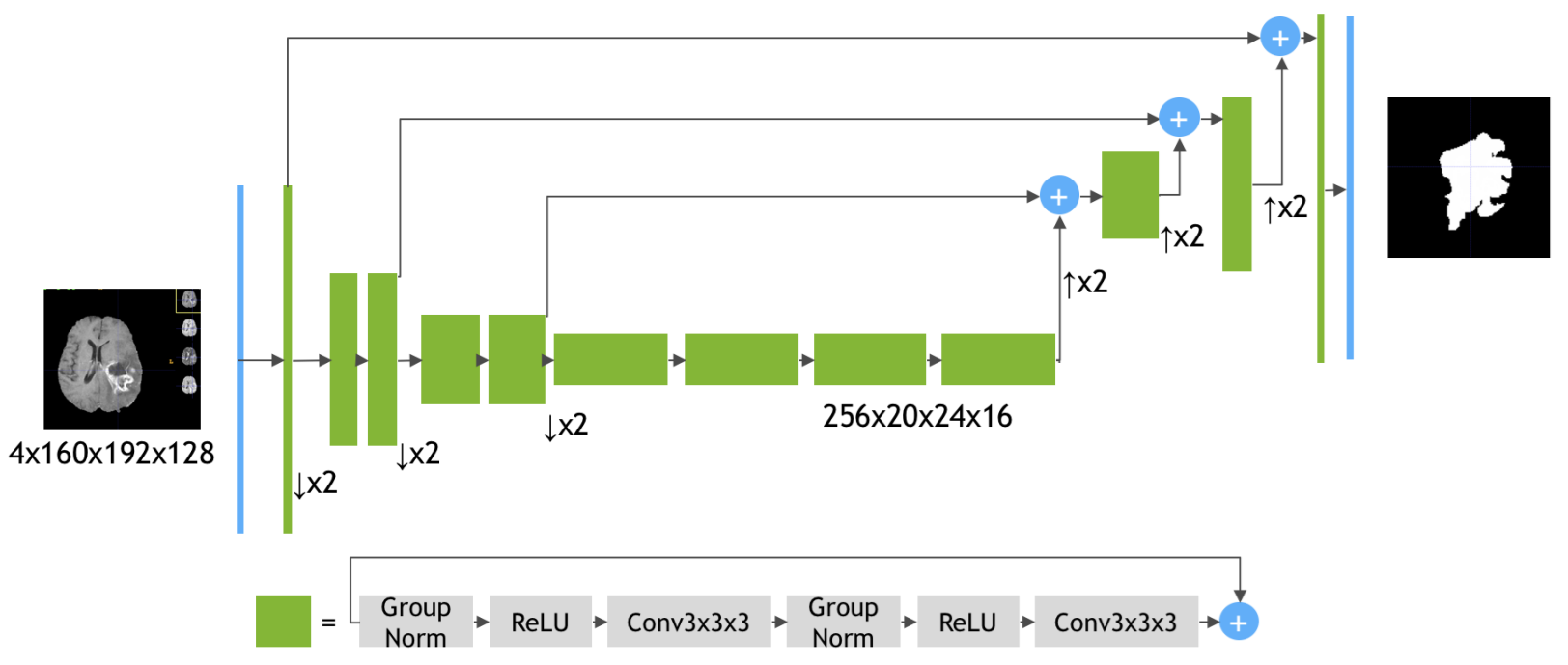}}

\caption{The SegResNet network \cite{myronenko20193d}.}
\label{fig:model2}
\end{figure}

We conducted a comprehensive performance analysis by comparing 2 different segmentation models with varying sets of hyper-parameters. Initially, we employed MedNeXt \cite{roy2023mednext}, a novel 3D segmentation network inspired by ConvNeXt architecture, recently introduced into the field. Alongside this, we utilized SegResNet \cite{myronenko20193d}, a CNN-based segmentation model developed by the winning team of the BraTs 2018 challenge. These models were trained to predict 3 classes in 3 different output channels (TC, WT, ET); however, we also conducted experiments where we trained the models separately for one underperforming class (ET). In addition, our model input size is $128\times128\times128$. The models are illustrated in Figure \ref{fig:model} and Figure \ref{fig:model2}. The details are mentioned in the sections below.

\noindent \textbf{MedNeXt.}
architecture draws inspiration from vision transformer \cite{dosovitskiy2020image} and incorporates them into the kernel segmentation network design. This combines the benefits of ConvNeXT \cite{liu2022convnet}-like structures in a UNet \cite{ronneberger2015u}-like design. Consequently, MedNeXt harnesses the inherent strengths of CNN models while integrating transformer-inspired ConvNeXt blocks tailored for 3D segmentation tasks. Notably, this design also implements deep supervision (DS) that can alleviate the problem of vanishing gradients, thus enhancing model training. Our experimentation encompassed two variants of the MedNeXt model: 'base' (B) and 'medium' (M) from the standard MedNeXt implementation \footnote{https://github.com/MIC-DKFZ/MedNeXt}.

\noindent \textbf{SegResNet.}
adopts a CNN-based encoder-decoder architecture that exhibits a relatively straightforward yet highly effective design for the 3D segmentation task. This architecture was originally introduced by the BraTs 2018 winning team, who achieved the highest dice score in segmenting tumor sub-regions by employing an ensemble of ten models. SegResNet has a ResNet-based \cite{he2016identity} asymmetric architecture, containing UNet-like encoder-decoder blocks but with skip connections on both the encoder and decoder, allowing better gradient propagation. It also uses Group Normalization \cite{wu2018group} that is suggested to work better by the authors, especially in small batch size scenarios.

Moreover, SegResNet incorporates an additional VAE (Variational Autoencoder) branch in the decoder during the training only. This VAE branch allows reconstructing the original image using features derived from the encoder bottleneck and does not contain skip connections from the encoder. This ensures a better regularization for the model training and allows the model to learn rich features for segmentation. During our experimentation, we conducted training iterations of both \textit{with} and \textit{without} the VAE regularization in order to assess its impact on the performance across the targeted tasks.

\subsection{Inference}

\noindent \textbf{Prediction from a single network.}
Our model input size is $128\times128\times128$, smaller than the MR image size. We implement the sliding window inference technique with 0.5 overlaps to predict tumor probabilities for each voxel. We apply test-time-augmentation (TTA) by flipping an input image through all possible flip combinations (8 combinations) and aggregating the mean probabilities.

\noindent \textbf{Ensemble.}
We train each model on the 5-fold CV setting, resulting in 5 trained networks for every training. To predict an input image during inference, we pass the image to each network to estimate tumor probabilities. Then, we aggregate the outputs from the 5-fold CV networks by taking the mean probabilities.

Ensembling multiple models can help improve overall performance \cite{brats22winner} by leveraging the inherent strength of every model. In this work, we ensemble models output on probability level by weighted averaging to give importance to every model on each channel $(weight\_tc, weight\_wt, weight\_et)$. The pseudocode of our model ensembling is given in Algorithm \ref{alg:model_ensembling}.

\begin{algorithm}
    \caption{Model Ensembling}\label{alg:model_ensembling}
    \begin{algorithmic}
        \Require $N$ models with each corresponding weighting for every channel (TC, WT, ET) and an input brain MRI scan $x \in \mathbb{R}^{3\times H \ \times W \times D}$
        
        \State $y \gets \mathbf{0}^{3 \times H \times W \times D}$
        \State $sum_w \gets \mathbf{0}^{3}$

        \For{$n=1,2,\dots N$}
            \State $y \gets y + \texttt{models}[n](x) * \texttt{weightings}[n]$
            \State $sum_w \gets sum_w + \texttt{weightings}[n]$
        \EndFor

        \State $y \gets y / sum_w$

        \State \Return y
    \end{algorithmic}
\end{algorithm}

\noindent \textbf{Postprocessing.} \label{sec:postprocessing} The postprocessing step plays a crucial role in the overall performance, especially for this year’s competition, as the organizer decided to change the evaluation focus from study-wise to lesion-wise performance, where False positive (FP) and negative (FN) are penalized severely with 0.0 Dice and 374 HD95 scores. Our experiments show that raw segmentation prediction contains many FPs due to small-size predicted lesions. To alleviate this, we do the following for each output channel (TC, WT, and ET); \begin{enumerate}
    \item Perform thresholding with a specific threshold for each channel.
    \item Perform connected component analysis to group predicted connected tumor voxels into lesions.
    \item Filter every group based on tumorous voxel count and the mean of tumorous voxel probabilities.
\end{enumerate}
In short, we implement two postprocessing functions as described in Algorithm \ref{alg:as_discrete} and Algorithm \ref{alg:filter_objects}.

\begin{algorithm}
    \caption{AsDiscrete($T_{TC}, T_{WT}, T_{ET}$)}\label{alg:as_discrete}
    \begin{algorithmic}
        \Require Threshold values for each channel ($T_{TC}, T_{WT}, T_{ET}$), and a predicted tumor heatmap $x \in \mathbb{R}^{3\times H \ \times W \times D}$ where the channels correspond to TC, WT, and ET respectively
        \Ensure $0<T_{TC}, T_{WT}, T_{ET}<1$
        
        \State $y \gets \mathbf{0}^{3\times H \ \times W \times D}$
        \For{$w, h, d = \texttt{range}(W), \texttt{range}(H), \texttt{range}(D)$}
            \If{$x[1,w,h,d] \geq T_{TC}$}
                \State $y[1,w,h,d] \gets 1$
            \EndIf
            \If{$x[2,w,h,d] \geq T_{WT}$}
                \State $y[2,w,h,d] \gets 1$
            \EndIf
            \If{$x[3,w,h,d] \geq T_{ET}$}
                \State $y[3,w,h,d] \gets 1$
            \EndIf
        \EndFor
    \State \Return $y$
    \end{algorithmic}
\end{algorithm}

\begin{algorithm}
    \caption{FilterObjects($T_{s,u}$, $T_{s,l}$, $T_{p,u}$, $T_{p,m}$)}\label{alg:filter_objects}
    \begin{algorithmic}
    \Require A predicted tumor heatmap and binary map $x_p, x_b \in \mathbb{R}^{H \ \times W \times D}$, upper size threshold ($T_{s,u}$), lower size threshold ($T_{s,l}$), upper probability threshold ($T_{p,u}$), and mid probability threshold ($T_{p,m}$)

    \Ensure $T_{s,u} \geq T_{s,l}$, $T_{s,l} \geq 0$, and $0 \leq T_{p,u}, T_{p,m} < 0$
    
    \State $y \gets \mathbf{0}^{H \ \times W \times D}$
    \State $y_{cc} \gets \texttt{get\_connected\_components}(x_b)$
    \State $N_{cc} \gets \texttt{get\_the\_number\_of\_ccs}(y_{cc})$
    \For{$n \in \texttt{range}(N_{cc})$}
        \State $\texttt{size} \gets \texttt{count\_tumor\_pixels\_of\_nth\_cc}(y_{cc}, n)$
        \State $\texttt{mean} \gets \texttt{get\_mean\_prob\_of\_nth\_cc}(x_p, y_{cc}, n)$
        \If{$\texttt{size} \geq T_{s,u}$}
            \If{$\texttt{mean} \geq T_{p,u}$}
                $y \gets \texttt{insert\_cc\_to\_y}(y, y_{cc}, n)$
            \EndIf
        \ElsIf{$T_{s,l} \leq \texttt{size} < T_{s,u}$}
            \If{$\texttt{mean} \geq T_{p,m}$}
                $y \gets \texttt{insert\_cc\_to\_y}(y, y_{cc}, n)$
            \EndIf
        \EndIf
    \EndFor
    \State \Return $y$
    \end{algorithmic}
\end{algorithm}

\subsection{Experimental setup}
We follow the 5-fold CV training setting by partitioning the training data into five subsets, performing training on four of them, and validating one subset on each iteration. We train our networks based on the region-based training mechanism \cite{nnunetforbrats} for 150 epochs, batch size of 2, and apply on-the-fly data augmentation consisting of the random spatial crop to $128\times128\times128$ size, random flips, and random intensity scaling as well as shifting. For the objective function, we apply batch dice loss and focal loss with 2.0 $\gamma$ and then sum them to get the total loss. We optimize our networks using AdamW optimizer \cite{DBLP:conf/iclr/LoshchilovH19}, and we use the cosine-annealing with linear-warmup scheduler. The optimizer and scheduler hyperparameters are 1e-4 base learning rate (LR), 1e-6 weight decay, 8 warmup epochs, 1e-7 initial LR, 1e-6 final LR, and 150 maximum epochs. In addition, we conducted some experiments involving deep supervision by additionally applying the loss at each decoder stage and giving lower importance weights on lower resolutions by a factor of 1/2.

We apply \textit{AsDiscrete} (Algorithm \ref{alg:as_discrete}) and \textit{FilterObjects} (Algorithm \ref{alg:filter_objects}) on each channel at the output prediction. The postprocessing hyperparameters are selected experimentally as we found that different model configurations have different output characteristics, leading to different suboptimum hyperparameters.



\section{Results}

\begin{figure}[t!]
\centering
{\includegraphics[width=0.85\columnwidth]{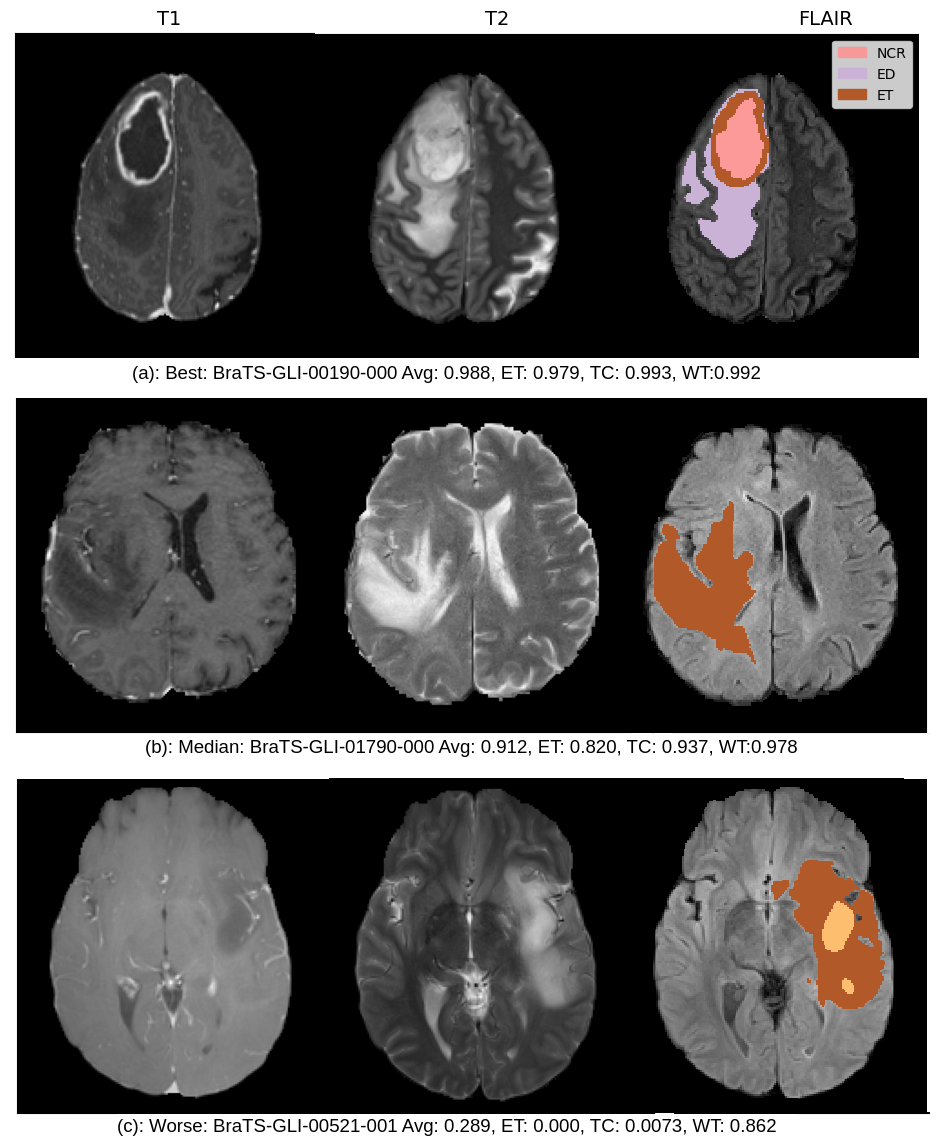}}

\caption{The figure shows the qualitative results for the adult-glioma task on the validation sample for three cases.}
\label{fig:qual_adults}
\end{figure}

{
\begin{table}
\centering
\caption{
Ablation study on post-processing using 5-fold CV. We utilize MedNeXt B-3 with deep supervision. Our post-processing steps significantly affect the BraTS 2023 Score, while the Legacy Score is not much affected. \textbf{Notes}: (a) Test-time augmentations (TTA), (b) Replace ET to TC if total predicted ET area is small, (c) Filter connected components (tumor objects) based on size, (d) Filter connected components based on mean confidence of each tumor object.
}
\tiny
\begin{tabular}{cccc|rrrr|rrrr|rrrr|rrrr}
\rowcolor{TableGray} & & & & \multicolumn{8}{|c}{\textbf{BraTS 2023 Score}} & \multicolumn{8}{|c}{\textbf{Legacy Score}} \\
\rowcolor{TableGray} & & & & \multicolumn{4}{|c|}{Dice} & \multicolumn{4}{c|}{HD95} & \multicolumn{4}{c}{Dice} & \multicolumn{4}{|c}{HD95} \\
\rowcolor{TableGray} \multirow{-3}{*}{a} & \multirow{-3}{*}{b} & \multirow{-3}{*}{c} & \multirow{-3}{*}{d} & \multicolumn{1}{|c}{ET} & \multicolumn{1}{c}{TC} & \multicolumn{1}{c}{WT} & \multicolumn{1}{c|}{Avg} & \multicolumn{1}{c}{ET} & \multicolumn{1}{c}{TC} & \multicolumn{1}{c}{WT} & \multicolumn{1}{c}{Avg} & \multicolumn{1}{|c}{ET} & \multicolumn{1}{c}{TC} & \multicolumn{1}{c}{WT} & \multicolumn{1}{c}{Avg} & \multicolumn{1}{|c}{ET} & \multicolumn{1}{c}{TC} & \multicolumn{1}{c}{WT} & \multicolumn{1}{c}{Avg} \\
\hline
 & & & & 78.82 & 84.55 & 70.88 & 78.08 & 47.66 & 33.84 & 93.57 & 58.36 & 87.14 & 91.38 & 93.29 & 90.60 & 11.69 & 7.50 & 6.59 & 8.59 \\
\checkmark & & & & 79.99 & 86.30 & 77.45 & 81.25 & 43.25 & 26.62 & 67.94 & 45.94 & 87.24 & 91.40 & 93.46 & 90.70 & 11.82 & \textbf{7.19} & 6.54 & 8.52 \\
\checkmark & \checkmark & & & 80.91 & 86.30 & 77.45 & 81.55 & 41.26 & 26.62 & 67.94 & 45.27 & 88.08 & 91.40 & 93.40 & 90.98 & 11.06 & \textbf{7.19} & 6.54 & 8.26 \\
\checkmark & & \checkmark & & 85.51 & 88.63 & 89.06 & 87.74 & 21.83 & 19.28 & 22.96 & 21.36 & 88.29 & 91.10 & 93.34 & 90.91 & 10.49 & 9.89 & 7.02 & 9.13 \\
\checkmark & & & \checkmark & 84.94 & 88.42 & \textbf{89.61} & 87.66 & 24.47 & \textbf{18.36} & \textbf{20.83} & 21.22 & 88.19 & \textbf{91.46} & \textbf{93.48} & \textbf{91.04} & \textbf{10.34} & 7.47 & \textbf{6.35} & \textbf{8.06} \\
\rowcolor{TableLightCyan} \checkmark & & \checkmark & \checkmark & \textbf{86.01} & \textbf{88.70} & 89.44 & \textbf{88.05} & \textbf{20.59} & {19.11} & 21.59 & \textbf{20.43} & \textbf{88.38} & 91.07 & 93.36 & 90.94 & 10.35 & 10.10 & 6.81 & 9.09
\end{tabular}
\label{tab:5-cv-glioma}
\end{table}
}

\begin{table}
\centering
\caption{Adult-glioma performance on the validation leaderboard. DS indicates using Deep Supervision during training. The postprocessing hyperparameters for each submission were selected experimentally. The final postprocessing steps (*) are {\small AsDiscrete(0.5, 0.5, 0.4), FilterObjects(2000, 100, 0.85, 0.925) for WT, FilterObjects(95, 70, 0.71, 0.5) for ET, FilterObjects(350, 350, 0, 0) for TC}. The final ensemble weightings are MedNeXt-DS=$(0, 1, 1)$, SegResNet-DS=$(0, 1, 0)$, SegResNet=$(1, 0, 0)$.
}
\begin{tabular}{l|rrrr|rrrr}
\rowcolor{TableGray} & \multicolumn{4}{c|}{Dice} & \multicolumn{4}{c}{HD95} \\
\cline{2-9}
\rowcolor{TableGray} \multirow{-2}{*}{Model} & \multicolumn{1}{c}{ET} & \multicolumn{1}{c}{TC} & \multicolumn{1}{c}{WT} & \multicolumn{1}{c|}{Avg} & \multicolumn{1}{c}{ET} & \multicolumn{1}{c}{TC} & \multicolumn{1}{c}{WT} & \multicolumn{1}{c}{Avg} \\
\hline
SegResNet & 0.8280 & 0.8606 & 0.9044 & 0.8643 & 23.90 & 17.89 & 12.52 & 18.10 \\
SegResNet-DS & 0.8239 & 0.8595 & 0.9016 & 0.8617 & 27.46 & 16.24 & 13.48 & 19.06 \\
MedNeXt & 0.8400 & 0.8486 & 0.9059 & 0.8648 & 22.25 & 26.71 & 12.49 & 20.48 \\
MedNeXt-DS & 0.8363 & 0.8486 & 0.9051 & 0.8633 & 23.92 & 28.27 & 12.65 & 21.61 \\
\begin{tabular}[c]{@{}l@{}}MedNeXt-DS \\ \,\,\,\,+ SegResNet-DS\end{tabular} & 0.8346 & 0.8622 & \textbf{0.9063} & 0.8677 & 23.70 & 18.71 & \textbf{11.70} & 18.04 \\
\rowcolor{TableLightCyan} \begin{tabular}[c]{@{}l@{}}
*MedNeXt-DS \\ \,\,\,\,+ SegResNet-DS \\ \,\,\,\,+ SegResNet\end{tabular} & \textbf{0.8432} & \textbf{0.8627} & \textbf{0.9063} & \textbf{0.8707} & \textbf{17.37} & \textbf{13.10} & \textbf{11.70} & \textbf{14.06}
\end{tabular}
\label{tab:1}
\end{table}


\begin{table}
\centering
\caption{Performance on the test set in both tasks. Our final submission for the adult glioma segmentation task was ranked $3^{rd}$ in the final test set leaderboard. DS indicates using Deep Supervision during training, and ET indicates the models specifically trained for class ET in the pediatric tumor segmentation task.}
\begin{tabular}{l|l|rrrr|rrrr}
\rowcolor{TableGray} & & \multicolumn{4}{c|}{Dice} & \multicolumn{4}{c}{HD95} \\
\cline{3-10}
\rowcolor{TableGray} \multirow{-2}{*}{Task} & \multirow{-2}{*}{Model} & \multicolumn{1}{c}{ET} & \multicolumn{1}{c}{TC} & \multicolumn{1}{c}{WT} & \multicolumn{1}{c|}{Avg} & \multicolumn{1}{c}{ET} & \multicolumn{1}{c}{TC} & \multicolumn{1}{c}{WT} & \multicolumn{1}{c}{Avg} \\
\hline

\begin{tabular}[c]{@{}l@{}}Adult\\ \,\,\,\,Glioma\end{tabular} &
\begin{tabular}[c]{@{}l@{}}MedNeXt-DS +\\ \,\,\,\, SegResNet-DS\end{tabular} & 0.8198 & 0.8233 & 0.8508 & 0.8313 & 35.15 & 39.86 & 34.12 & 36.38 \\
\hline
\begin{tabular}[c]{@{}l@{}}Pediatric\\ \,\,\,\,Tumors\end{tabular} &
\begin{tabular}[c]{@{}l@{}}SegResNet-ET +\\ \,\,\,\,SegResNet\end{tabular} & 0.5522 & 0.77 & 0.7755 & 0.6992 & 45.32 & 30.35 & 30.45 & 35.37 \\


\end{tabular}
\label{tab:testset_performance}
\end{table}


\begin{table}
\caption{The table presents the validation results for BraTs-Pediatrics segmentation obtained from the leaderboard. In the WT (weighted-loss) training scheme, the focal loss weight is set to 2. Our final prediction is the result of various model combinations. Models specifically trained for class ET are marked as **-ET. The optimal performance was attained with a 5-Fold-CV of SegResNet (predicting WT and TC) and an additional 5-Fold-CV of SegResNet (predicting ET).}

    \centering
    \begin{tabular}{l|llll|llll}
\rowcolor{TableGray} & \multicolumn{4}{c|}{Dice} & \multicolumn{4}{c}{HD95} \\
\cline{2-9}
\rowcolor{TableGray} \multirow{-2}{*}{Model} & \multicolumn{1}{c}{ET} & \multicolumn{1}{c}{TC} & \multicolumn{1}{c}{WT} & \multicolumn{1}{c|}{Avg} & \multicolumn{1}{c}{ET} & \multicolumn{1}{c}{TC} & \multicolumn{1}{c}{WT} & \multicolumn{1}{c}{Avg} \\
\hline
            
        MedNeXt & 0.3338 & 0.7829 & 0.8291 & 0.6586 & 202.33 & \textbf{16.76} & 20.96 & 60.17 \\ 
        SegResNet & 0.2674 & \textbf{0.7930} & \textbf{0.8334} & 0.6313 & 183.82 & 19.98 & \textbf{19.32} & 55.94 \\         
        SegResNet-WL & 0.3238 & 0.7588 & 0.8060 & 0.6295 & 204.02 & 24.75 & 26.18 & 63.90 \\ 
        \shortstack{SegFormer ET\\+ MedNeXt}  & 0.5015 & 0.7829 & 0.8291 & 0.7045 & 149.81 & 16.76 & 20.97 & 47.06 \\
        \shortstack{MedNeXt ET model\\+ MedNeXt} & 0.4004 & 0.7765 & 0.8325 & 0.6698 & 189.98 & 17.14 & 19.61 & 56.85 \\ 
        \shortstack {SegResNet ET model + \\SegResNet-VAE} & 0.5595 & 0.7777 & 0.8018 & 0.7130 & 124.91 & 21.56 & 29.76 & \textbf{44.23} \\ 
        \rowcolor{TableLightCyan} \shortstack {SegResNet ET model \\ + SegResNet} & \textbf{0.5595} & 0.78106 & 0.8206 & \textbf{0.7204} & \textbf{124.91} & 25.30 & 24.77 & 58.32 \\
    \end{tabular}
\label{tab:2}
\end{table}

We trained our pipeline using the 5-fold cross-validation (CV) on the training data and performed an evaluation using the internal validation set to select the best set of hyperparameters. We further utilized highly effective post-processing steps (see section \ref{sec:postprocessing}) on the model predictions to get the final output, which was submitted to the online leaderboard. After rigorous experimentation on different models on the internal validation set, we narrowed it down to two models for the external validation based on the online leaderboard. SegResNet and MedNeXt models are selected for the final submissions. The dice similarity coefficient (DSC) and 95\% Hausdorff distance (HD95) from the online validation for the Adults-Glioma and Pediatrics task are mentioned in Table \ref{tab:1} and Table \ref{tab:2},  respectively. Furthermore, the performance for both tasks on the final test set is given in Table \ref{tab:testset_performance}.



Table \ref{tab:5-cv-glioma} shows the effect of different post-processing steps on the performance evaluated using the local validation set. This ablations study used the MedNeXt-B-3 model with deep supervision (DS). The best-performing setting is achieved by using test-time augmentations (TTA) along with filtering connected components based on each tumor object's size and mean confidence. We achieved approximately a 10\% increase in average Dice score using and a large drop in the HD95 distance, suggesting the effectiveness of the used post-processing.

We further report the performance metrics on the adult Glioma task in table~\ref{tab:1}. In the vanilla 5-fold CV, both baseline models (SegResNet and MedNeXt) achieve a similar performance of approx. 0.86 mean DSC. Moreover, using DS shows a slight positive trend in HD95 distance. Following this, we create an ensemble of MedNeXt-DS, SegResNet-DS, and SegResNet without DS to achieve the highest performance in terms of mean DSC and HD95, 0.871 and 14.06, respectively. Our best-performing setting is a weighted combination of all three used models, where we used MedNeXt-DS for class TC and WT, SegResNet-DS for class TC, and SegResNet for ET class only. We used this combination for the final submission. Table \ref{tab:testset_performance} shows the performance on the hidden test set, where our approach achieved 0.8313 and 36.38 mean DSC and HD95, respectively. The qualitative results for the best, median, and worst performing validation samples are shown in Figure \ref{fig:qual_adults}.

\begin{figure}[t!]
\centering
{\includegraphics[width=0.85\columnwidth]{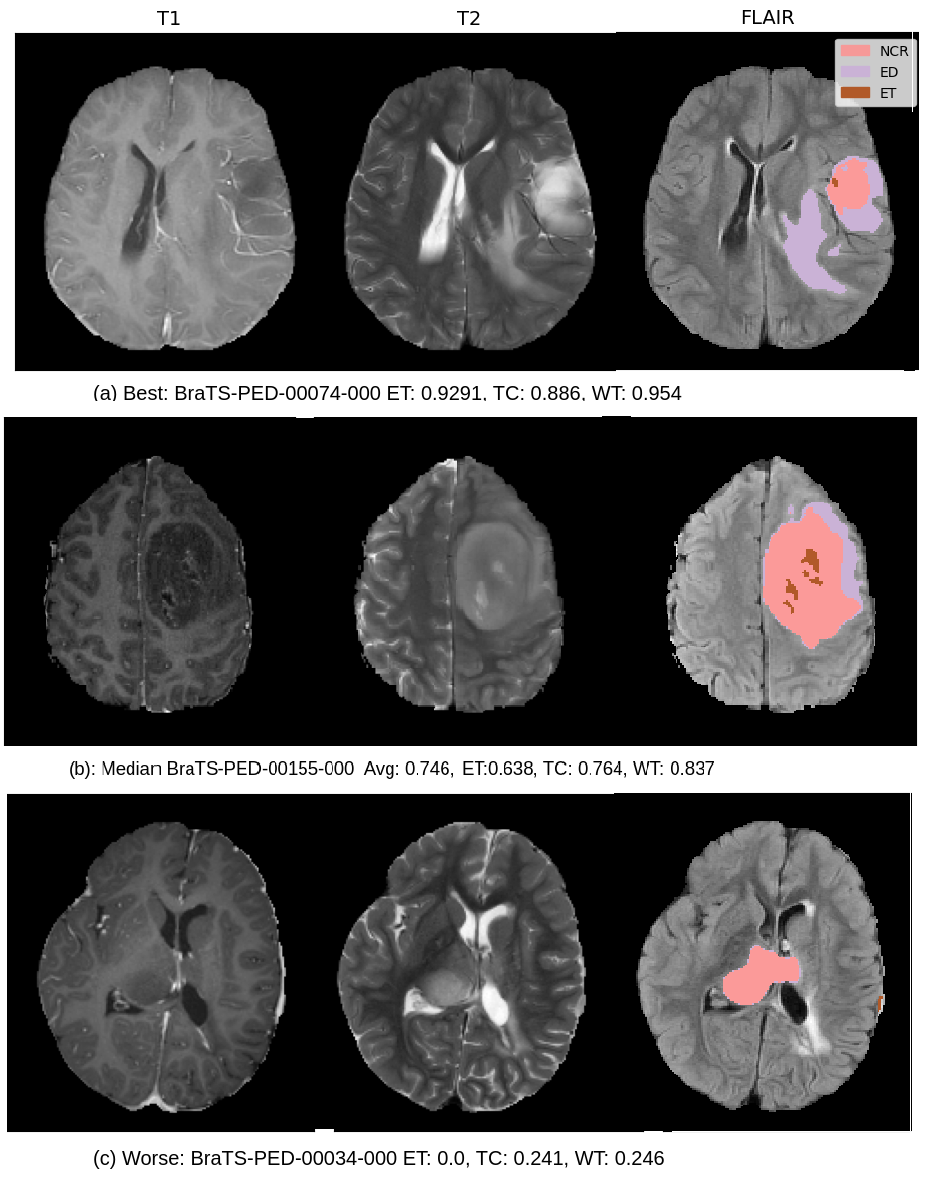}}

\caption{The figure shows the qualitative results for the pediatric task on the validation sample for the best-performing model.}
\label{fig:qual_peds}
\end{figure}

Following a similar experimental setting, we applied our proposed pipeline to the pediatric tumor segmentation task as well. Table \ref{tab:2} shows that the baseline MedNeXt and SegResNet models achieve a mean DSC of 0.65 and 0.63 and a mean HD95 of 60.17 and 55.94, respectively, with sub-optimal performance for the ET class. We developed two techniques to tackle this issue: (i) a weighted loss (WL) based model that penalizes the ET class more than other classes and (ii) a standalone model for the ET class only. While SegResNet-WL showed some improvement in the ET class performance, it did not perform on par with approach (ii), where a separate network is trained for the ET class only. Our best-performing model for this task is a combination of a SegResNet-ET and a multi-class SegResNet model, achieving a mean DSC of 0.72 and HD95 of 58.32. We used our best-performing combination for the test set submission. Table \ref{tab:testset_performance} shows that our approach achieved 0.6992 and 35.37 mean DSC and HD95 on the hidden test set, respectively. The qualitative results for the best, median, and worst performing validation samples are shown in figure~\ref{fig:qual_peds}.




\section{Discussion}
In this paper, we summarize our proposed methodology for the BraTS 2023 adult and pediatrics glioma segmentation competition. Our pipeline is based on the two highly efficient segmentation models, namely; MedNeXt and SegResNet. Our work benefited from the power of DS and an ensemble strategy involving heterogeneous models to tackle a challenging real-life problem. The primary focus was based on integrating suitable post-processing steps with deep supervision and the ensemble of diverse models, resulting in substantial performance gains. We conducted comprehensive experiments to show the significance of understanding the clinical problem and implementing domain-specific processing steps to augment the efficiency of deep-learning models, ultimately providing a significant boost in performance. The post-processing steps stood very helpful in following this year's new scoring system that is designed carefully based on the clinical diagnosis and heavily penalizes missing even a small tumor region. 

Furthermore, we used a similar approach for both tasks, motivated by the similarity between both tasks; however, the performance in both tasks remained significantly different. In the adult-glioma challenge, all three classes showed similar results, while the ET class in the pediatric dataset suffered much lower performance when compared to the other two classes. This discrepancy can potentially be due to the size of the ET regions as compared to the entire brain volume and other classes. The models find it hard to distinguish the regions, especially when one class is imbalanced, as in the case of the pediatric dataset. Another reason is that the ET class is a small area within the TC region, which itself is within the WT region. This suppresses the ET class significantly, affecting smooth model learning. Furthermore, the size of the available dataset for the pediatric task is significantly smaller than the adult-glioma dataset, which could also contribute to this performance difference in both tasks. 


\section{Conclusion}
The paper studies different approaches to segmenting the tumor region in the brain. We used the BraTS 2023 challenge datasets for the adult glioma and pediatric tumor. Both datasets are multi-modal, multi-class segmentation tasks, with four modalities to input and three classes to predict. The automatic approach for segmenting tumor regions is highly beneficial for clinical practice, helping clinicians speed up their work. Our approach combines the advantages of deep supervision and an ensemble of models for the successful segmentation of the brain tumor from the MR images. To conclude, we achieve third place in the \texttt{BraTS 2023 Adult Glioma Challenge.}


\newpage

\bibliography{biblography}

\begin{thebibliography}{10}

\bibitem{alexander2017adult}
B.~M. Alexander and T.~F. Cloughesy.
\newblock Adult glioblastoma.
\newblock {\em Journal of Clinical Oncology}, 35(21):2402--2409, 2017.

\bibitem{baid2021rsna}
U.~Baid, S.~Ghodasara, S.~Mohan, M.~Bilello, E.~Calabrese, E.~Colak,
  K.~Farahani, J.~Kalpathy-Cramer, F.~C. Kitamura, S.~Pati, et~al.
\newblock The rsna-asnr-miccai brats 2021 benchmark on brain tumor segmentation
  and radiogenomic classification.
\newblock {\em arXiv preprint arXiv:2107.02314}, 2021.

\bibitem{bakas10segmentationgbm}
S.~Bakas, H.~Akbari, A.~Sotiras, M.~Bilello, M.~Rozycki, J.~Kirby, J.~Freymann,
  K.~Farahani, and C.~Davatzikos.
\newblock Segmentation labels and radiomic features for the pre-operative scans
  of the tcga-gbm collection (2017).
\newblock {\em DOI: https://doi. org/10.7937 K}, 9.

\bibitem{bakas2017segmentationlgg}
S.~Bakas, H.~Akbari, A.~Sotiras, M.~Bilello, M.~Rozycki, J.~Kirby, J.~Freymann,
  K.~Farahani, and C.~Davatzikos.
\newblock Segmentation labels and radiomic features for the pre-operative scans
  of the tcga-lgg collection.
\newblock {\em The cancer imaging archive}, 286, 2017.

\bibitem{bakas2017advancing}
S.~Bakas, H.~Akbari, A.~Sotiras, M.~Bilello, M.~Rozycki, J.~S. Kirby, J.~B.
  Freymann, K.~Farahani, and C.~Davatzikos.
\newblock Advancing the cancer genome atlas glioma mri collections with expert
  segmentation labels and radiomic features.
\newblock {\em Scientific data}, 4(1):1--13, 2017.

\bibitem{dosovitskiy2020image}
A.~Dosovitskiy, L.~Beyer, A.~Kolesnikov, D.~Weissenborn, X.~Zhai,
  T.~Unterthiner, M.~Dehghani, M.~Minderer, G.~Heigold, S.~Gelly, et~al.
\newblock An image is worth 16x16 words: Transformers for image recognition at
  scale.
\newblock {\em arXiv preprint arXiv:2010.11929}, 2020.

\bibitem{he2016identity}
K.~He, X.~Zhang, S.~Ren, and J.~Sun.
\newblock Identity mappings in deep residual networks.
\newblock In {\em Computer Vision--ECCV 2016: 14th European Conference,
  Amsterdam, The Netherlands, October 11--14, 2016, Proceedings, Part IV 14},
  pages 630--645. Springer, 2016.

\bibitem{nnunetforbrats}
F.~Isensee, P.~F. J{\"a}ger, P.~M. Full, P.~Vollmuth, and K.~H. Maier-Hein.
\newblock nnu-net for brain tumor segmentation.
\newblock In A.~Crimi and S.~Bakas, editors, {\em Brainlesion: Glioma, Multiple
  Sclerosis, Stroke and Traumatic Brain Injuries}, pages 118--132, Cham, 2021.
  Springer International Publishing.

\bibitem{karargyris2023federated}
A.~Karargyris, R.~Umeton, M.~J. Sheller, et~al.
\newblock Federated benchmarking of medical artificial intelligence with
  medperf.
\newblock {\em Nature Machine Intelligence}, 5:799--810, 2023.

\bibitem{kazerooni2023brain}
A.~F. Kazerooni et~al.
\newblock The brain tumor segmentation ({BraTS}) challenge 2023: Focus on
  pediatrics ({CBTN-CONNECT-DIPGR-ASNR-MICCAI BraTS-PEDs}).
\newblock {\em arXiv preprint arXiv:2305.17033}, 2023.

\bibitem{liu2022convnet}
Z.~Liu, H.~Mao, C.-Y. Wu, C.~Feichtenhofer, T.~Darrell, and S.~Xie.
\newblock A convnet for the 2020s.
\newblock In {\em Proceedings of the IEEE/CVF conference on computer vision and
  pattern recognition}, pages 11976--11986, 2022.

\bibitem{DBLP:conf/iclr/LoshchilovH19}
I.~Loshchilov and F.~Hutter.
\newblock Decoupled weight decay regularization.
\newblock In {\em 7th International Conference on Learning Representations,
  {ICLR} 2019, New Orleans, LA, USA, May 6-9, 2019}. OpenReview.net, 2019.

\bibitem{menze2014multimodal}
B.~H. Menze, A.~Jakab, S.~Bauer, J.~Kalpathy-Cramer, K.~Farahani, J.~Kirby,
  Y.~Burren, N.~Porz, J.~Slotboom, R.~Wiest, et~al.
\newblock The multimodal brain tumor image segmentation benchmark (brats).
\newblock {\em IEEE transactions on medical imaging}, 34(10):1993--2024, 2014.

\bibitem{myronenko20193d}
A.~Myronenko.
\newblock 3d mri brain tumor segmentation using autoencoder regularization.
\newblock In {\em Brainlesion: Glioma, Multiple Sclerosis, Stroke and Traumatic
  Brain Injuries: 4th International Workshop, BrainLes 2018, Held in
  Conjunction with MICCAI 2018, Granada, Spain, September 16, 2018, Revised
  Selected Papers, Part II}, volume~4. Springer International Publishing, 2019.

\bibitem{owrangi2018mri}
A.~M. Owrangi, P.~B. Greer, and C.~K. Glide-Hurst.
\newblock Mri-only treatment planning: benefits and challenges.
\newblock {\em Physics in Medicine \& Biology}, 63(5):05TR01, 2018.

\bibitem{pereira2016brain}
S.~Pereira, A.~Pinto, V.~Alves, and C.~A. Silva.
\newblock Brain tumor segmentation using convolutional neural networks in mri
  images.
\newblock {\em IEEE transactions on medical imaging}, 35(5):1240--1251, 2016.

\bibitem{rohlfing2010sri24}
T.~Rohlfing et~al.
\newblock The {SRI24} multichannel atlas of normal adult human brain structure.
\newblock {\em Human Brain Mapping}, 31(5):798--819, 2010.

\bibitem{ronneberger2015u}
O.~Ronneberger, P.~Fischer, and T.~Brox.
\newblock U-net: Convolutional networks for biomedical image segmentation.
\newblock {\em Medical Image Computing and Computer-Assisted Intervention
  (MICCAI)}, 9351:234--241, 2015.

\bibitem{roy2023mednext}
S.~Roy, G.~Koehler, C.~Ulrich, M.~Baumgartner, J.~Petersen, F.~Isensee, P.~F.
  Jaeger, and K.~Maier-Hein.
\newblock Mednext: Transformer-driven scaling of convnets for medical image
  segmentation.
\newblock {\em arXiv preprint arXiv:2303.09975}, 2023.

\bibitem{wu2018group}
Y.~Wu and K.~He.
\newblock Group normalization.
\newblock In {\em Proceedings of the European conference on computer vision
  (ECCV)}, pages 3--19, 2018.

\bibitem{xie2005semi}
K.~Xie, J.~Yang, Z.~Zhang, and Y.~Zhu.
\newblock Semi-automated brain tumor and edema segmentation using mri.
\newblock {\em European journal of radiology}, 56(1):12--19, 2005.

\bibitem{brats22winner}
R.~A. Zeineldin, M.~E. Karar, O.~Burgert, and F.~Mathis-Ullrich.
\newblock Multimodal cnn networks for brain tumor segmentation in mri: A brats
  2022 challenge solution.
\newblock In S.~Bakas, A.~Crimi, U.~Baid, S.~Malec, M.~Pytlarz, B.~Baheti,
  M.~Zenk, and R.~Dorent, editors, {\em Brainlesion: Glioma, Multiple
  Sclerosis, Stroke and Traumatic Brain Injuries}, pages 127--137, Cham, 2023.
  Springer Nature Switzerland.

\end{thebibliography}
\bibliographystyle{abbrv}

\end{document}